\documentclass[letterpaper, 10 pt, conference]{ieeeconf}  
\IEEEoverridecommandlockouts                             
\usepackage{graphics} % for pdf, bitmapped graphics files
\usepackage{epsfig}   % for postscript graphics files
\usepackage{times}    % assumes new font selection scheme installed
\usepackage{amsmath}  % assumes amsmath package installed
\usepackage{amssymb}  % assumes amsmath package installed
\usepackage{mathtools}
\usepackage{gensymb}
\usepackage[export]{adjustbox}
\usepackage{color}
\usepackage{leftidx}
\usepackage{algorithm,algorithmic}
\usepackage{tablefootnote}
\usepackage{booktabs}
\usepackage{cite}[noadjust]
\usepackage{siunitx}
\usepackage{nomencl}
\usepackage{soul}
\makenomenclature

\newtheorem{theorem}{Theorem}
\newtheorem{remark}{Remark}
\newtheorem{assum}{Assumption}
\newtheorem{property}{Property}

\usepackage [latin1]{inputenc}

\DeclareMathOperator{\diag}{diag}

\title{\LARGE \bf
Impedance and Stability Targeted Adaptation for \\
Aerial Manipulator with Unknown Coupling Dynamics
}

\author{Amitabh Sharma$^{1}$, Saksham Gupta$^{1}$, Shivansh Pratap Singh$^{1}$, 
Rishabh Dev Yadav$^{2}$, Hongyu Song$^{2}$, \\ Wei Pan$^{2}$, Spandan Roy$^{1}$, 
Simone Baldi$^{3}$%
\thanks{This work was supported in part by ``Aerial Manipulation'' under IHFC grand project (GP/2021/DA/032), in part by ``Capacity building for human resource development in Unmanned Aircraft System (Drone and related Technology)'', MeiTY, India, in part by the Natural Science Foundation of China grants 62233004 and 62073074, and in part by Jiangsu Provincial Scientific Research Center of Applied Mathematics grant BK20233002.}
\thanks{$^{1}$ The authors are with the Robotics Research Center, 
International Institute of Information Technology Hyderabad (IIIT-H), India, email: 
{\tt\small \{amitabh.sharma, saksham.g, shivansh.singh\}@research.iiit.ac.in, spandan.roy@iiit.ac.in}}%
\thanks{$^{2}$ The authors are with the Department of Computer Science, The University of Manchester, United Kingdom, email: 
{\tt\small \{rishabh.yadav, hongyu.song-3\}@postgrad.manchester.ac.uk, wei.pan@manchester.ac.uk}}%
\thanks{$^{3}${The author is with the School of Mathematics, Southeast University, Nanjing,
China, email:
{\tt\small simonebaldi@seu.edu.cn}}}
}

\begin{document}
\maketitle
\thispagestyle{empty}
\pagestyle{empty}

\setlength{\belowcaptionskip}{-10pt}

%%%%%%%%%%%%%%%%%%%%%%%%%%%%%%%%%%%%%%%%%%%%%%%%%%%%%%%%%%%%%%%%%%%%%%%%%%%%%%%%
\begin{abstract}
Stable aerial manipulation during dynamic tasks such as object catching, perching, or contact with rigid surfaces necessarily requires compliant behavior, which is often achieved via impedance control. Successful manipulation depends on how effectively the impedance control can tackle the unavoidable coupling forces between the aerial vehicle and the manipulator. However, the existing impedance controllers for aerial manipulator either ignore these coupling forces (in partitioned system compliance methods) or require their precise knowledge (in complete system compliance methods). Unfortunately, such forces are very difficult to model, if at all possible. To solve this long-standing control challenge, we introduce an impedance controller for aerial manipulator which does not rely on a priori knowledge of the system dynamics and of the coupling forces. The impedance control design can address unknown coupling forces, along with system parametric uncertainties, via suitably designed adaptive laws. The closed-loop system stability is proved analytically and experimental results with a payload-catching scenario demonstrate significant improvements in overall stability and tracking over the state-of-the-art impedance controllers using either partitioned  or complete system compliance.

\end{abstract}

%%%%%%%%%%%%%%%%%%%%%%%%%%%%%%%%%%%%%%%%%%%%%%%%%%%%%%%%%%%%%%%%%%%%%%%%%%%%%%%%
\section{Introduction}
\label{sec:sec1}

Autonomous Aerial Manipulators (AAMs) have emerged as versatile platforms that combine the agility of aerial vehicles with the dexterity of robotic arms, enabling operations in challenging environments such as disaster sites, industrial facilities, etc.~\cite{kim_iros13, heredia_iros14}. %Among many possible applications However, the dynamic coupling between the aerial platform and the manipulator—coupled with environmental uncertainties—poses critical control challenges, 
During manipulation tasks involving active interactions with the environment like object catching, perching, or contact with rigid surfaces, an AAM system is required to attain compliant behavior for stability. Traditionally, impedance control strategies are employed to achieve compliance and they can be broadly classified as follows:
\begin{itemize}
\item \textbf{Complete System Compliance (CSC):} In these methods, the entire vehicle-manipulator system is modeled as a unified dynamic entity, ~\cite{Lippiello_12, ollero_15, Cataldi_16, heredia_iros14}.  
However, in order to enforce impedance behavior, CSC designs depend heavily on accurate models of the complete system, including the coupling forces between the aerial vehicle and the manipulator. As a result, these methods can be sensitive to uncertainties.
    \item \textbf{Partitioned System Compliance (PSC):} These techniques either provide compliance only to the aerial vehicle (cf. \cite{rashad_energy_2022, kim_iros13, Zhang_TASE_24, guanya,7989314}) considering the manipulator to be an inert entity solely used for force feedback, or provide (possibly adaptive) compliance only to the manipulator with separate control for the aerial vehicle (cf. \cite{suarez2018physical, tognon2019truly, wang_impact_2024}). While these methods simplify the control design, they end up ignoring the coupling forces between the aerial vehicle and the manipulator. % leading to suboptimal performance in dynamic tasks.
\end{itemize}
Coupling forces between the aerial vehicle and the manipulator are unavoidable during dynamic tasks. 
However, dynamic changes in the center of mass and mass/inertia distribution during manipulation activities are very difficult, if at all possible,
to model and the same holds for the coupling forces
between the aerial vehicle and the manipulator (cf. \cite[Ch. 5.3]{orsag2018aerial}). Possibly due to these challenges, recent research works have been focusing more on PSC methods as compared to CSC methods. However, although more challenging, CSC methods have more potential, provided that modeling imperfections can be handled. 

Unfortunately, to the best of the authors' knowledge, the present literature does not provide any impedance control solution for AAM systems which takes into account the effects of coupling forces without relying on their modeling accuracy. This work solves this control challenge with the following highlights:

\begin{itemize}
    \item We propose an adaptive impedance control providing compliance to both the aerial vehicle and the manipulator (unlike \cite{suarez2018physical, tognon2019truly, wang_impact_2024}). 
    \item The proposed design embeds an adaptive mechanism that allows to remove a priori knowledge of system dynamics, including a priori knowledge
of coupling forces (unlike \cite{Lippiello_12, ollero_15, Cataldi_16, heredia_iros14}). 
    
\end{itemize}

Closed-loop system stability is analyzed via Lyapunov theory. The effectiveness of the proposed controller is verified over state-of-the-art controllers via a payload catching experiment with a quadrotor-based AAM.

This paper is organized as follows: Section~\ref{sec:sec2} details the system dynamics and control problem formulation. Section~\ref{sec:sec3} presents the proposed adaptive impedance controller. Section~\ref{sec:sec4} describes the experimental results, and Section~\ref{sec:sec5} summarizes the present work and future direction.

%%%%%%%%%%%%%%%%%%%%%%%%%%%%%%%%%%%%%%%%%%%%%%%%%%%%%%%%%%%%%%%%%%%%%%%%%%%%%%%%
\section{System Dynamics and Control Problem Formulation}
\label{sec:sec2}
The following notations are used in this paper:  $|| (\cdot)||$ and $\lambda_{\min}(\cdot)$ denote 2-norm and minimum eigenvalue of $(\cdot)$, respectively; $\boldsymbol I$ denotes the identity matrix with appropriate dimension and $\diag\lbrace \cdot, \cdots, \cdot \rbrace$ denotes a diagonal matrix; $\boldsymbol{0}_{n}$ denotes the vector of size $n$ whose entries are all zero.

\begin{figure}[t]
\centering
\includegraphics[scale=0.32]{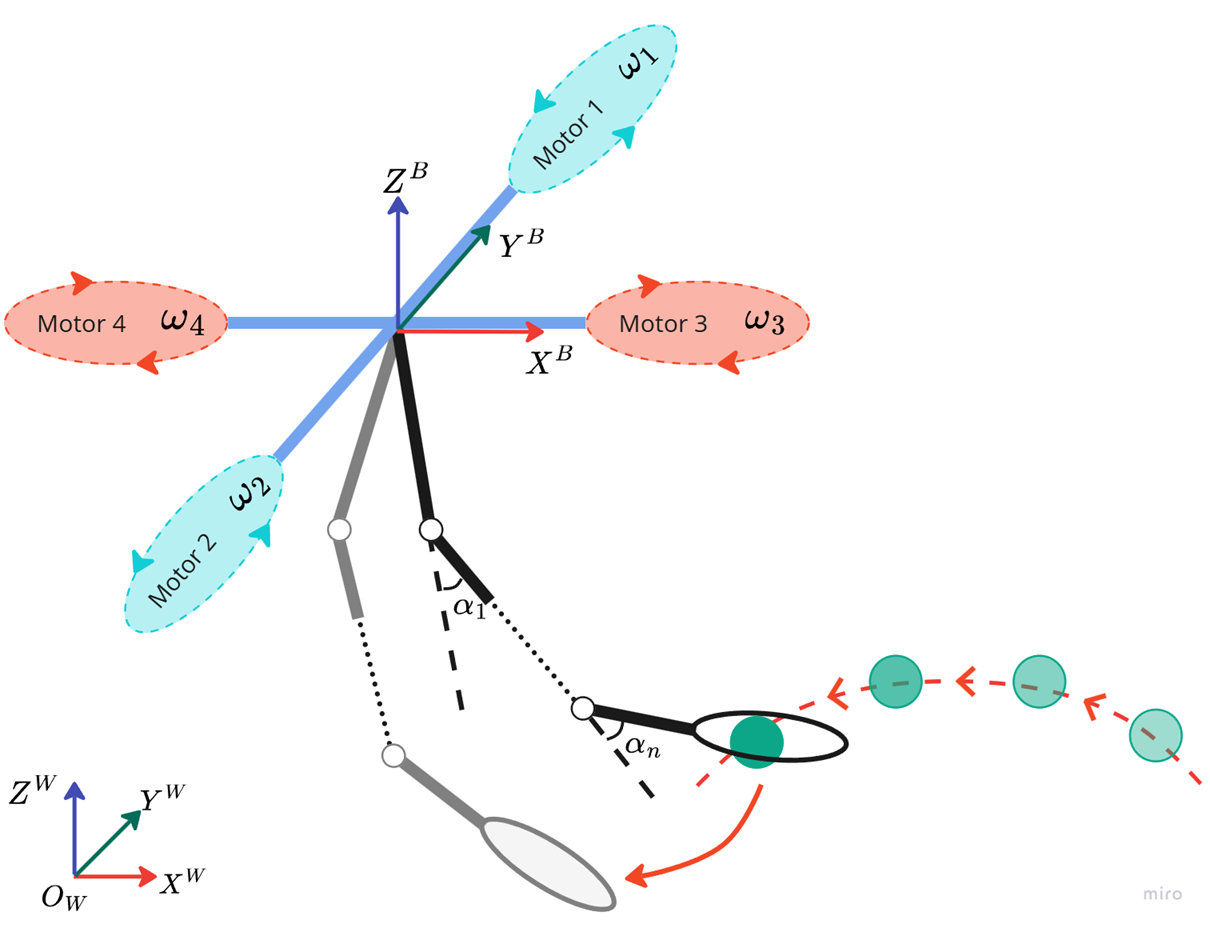}
\caption{A quadrotor-based AAM with an $n$-link manipulator catching an object (with relevant coordinate frames).}
\label{robot_model}
\end{figure}

\begin{table}[t]
\renewcommand{\arraystretch}{1.0}
\caption{Nomenclature}
\label{table_nomenclature}
\centering
\begin{tabular}{ll}
\toprule
$[\mathbf{X}^B \ \mathbf{Y}^B \ \mathbf{Z}^B ]$ & Quadrotor body-fixed frame \\
$[\mathbf{X}^W \ \mathbf{Y}^W \ \mathbf{Z}^W]$ & Earth-fixed frame \\
$\boldsymbol{R}_B^W \in \mathbb{R}^{3 \times 3}$ & $Z$-$Y$-$X$ Euler angle rotation matrix \\
$\mathbf{p} = [x \ y \ z]^{\top}$ & Quadrotor position in $[\mathbf{X}^W \ \mathbf{Y}^W \ \mathbf{Z}^W]$ \\
$\mathbf{q} = [\phi \ \theta \ \psi]^{\top}$ & Quadrotor roll, pitch, yaw \\
$\boldsymbol{\alpha} = [\alpha_1 \ \alpha_2 \ \ldots \ \alpha_n]^{\top}$ & Manipulator joint angles \\
$\mathbf{M}, \mathbf{C} \in \mathbb{R}^{(6+n) \times (6+n)}$ & Mass and Coriolis matrices \\
$\mathbf{g}, \mathbf{d} \in \mathbb{R}^{6+n}$ & Gravity vector and external disturbance\\
$\boldsymbol{\tau}_{p}, \boldsymbol{\tau}_{q} \in \mathbb{R}^{3}$ & Quadrotor position and attitude control inputs \\
$\boldsymbol{\tau}_{\alpha} \in \mathbb{R}^n$ & Joint control inputs of manipulator \\
$\mathbf{J}_\alpha \in \mathbb{R}^{3 \times n}$& Manipulator Jacobian in end-effector frame \\
$\mathbf{F}_{ext} \in \mathbb{R}^{3}$ & External force on end-effector \\
 $\mathbf{F}_{d} \in \mathbb{R}^{3}$&Desired force on end-effector\\

\bottomrule
\end{tabular}
\end{table}

The Euler-Lagrangian (EL) dynamical model a quadrotor-based AAM system with an $n$ degrees-of-freedom (DoF) manipulator system as in Fig. \ref{robot_model}, with symbols and system parameters as in Table \ref{table_nomenclature}, is given by \cite{arleo2013control}:
\begin{equation}\label{General_EL_dynamics}
\mathbf{M}(\boldsymbol{\chi})\,\ddot{\boldsymbol{\chi}}
+
\mathbf{C}(\boldsymbol{\chi},\dot{\boldsymbol{\chi}})\,\dot{\boldsymbol{\chi}}
+
\mathbf{g}(\boldsymbol{\chi})
+
\mathbf{d}
=
\boldsymbol{\tau}
+
\boldsymbol{\tau}_{\text{ext}},
\end{equation}
where $\boldsymbol{\chi} \triangleq  \begin{bmatrix} \boldsymbol{p}^\top ~~\boldsymbol{q}^\top ~~\boldsymbol{\alpha}^\top  \end{bmatrix}^\top $; 
$\boldsymbol{\tau} \triangleq  \begin{bmatrix} \boldsymbol{\tau}_p^\top ~~\boldsymbol{\tau}_q^\top ~~\boldsymbol{\tau}_\alpha^\top  \end{bmatrix}^\top $; 
$\boldsymbol{\tau}_{ext} \triangleq  \begin{bmatrix} \boldsymbol{0}_{3} ~~\boldsymbol{0}_3 ~~\mathbf{J}_\alpha^{\top} \mathbf{F}_{ext} \end{bmatrix}^\top $ denotes the external torques; 
$\boldsymbol{\tau}_{p} = \boldsymbol{R}^W_B \begin{bmatrix}
0 & 0 & u
\end{bmatrix}^\top$ is the generalized position control input with $u$ being total thrust. \emph{The coupling forces between the quadrotor and the manipulator are captured by the mass and Coriolis matrices in system dynamics (\ref{General_EL_dynamics}).}

In practice, such coupling forces are hard to model, resulting in large uncertainty in the structure and in the parameters of the mass and Coriolis matrices. This work addresses the modeling challenge by relaxing a priori knowledge on the system dynamics terms. We only make use of the following standard system properties that hold for the AAM owing to the EL mechanics \cite{spong2020robot, kim2013aerial}:
\begin{property}\label{prop_1}
$\mathbf{M}(\boldsymbol{\chi})$ is uniformly positive definite $\forall \boldsymbol{\chi}$ and 
$\exists\,\underline{m}, \overline{m}>0$ such that \ 
$0 < \underline{m}\,\mathbf{I} \,\le\, \mathbf{M}(\boldsymbol{\chi})\,\le\, \overline{m}\,\mathbf{I}.$
\end{property}

\begin{property}\label{prop_2}
$\exists\,\bar{c}, \bar{g}, \bar{d}>0$ such that 
$\|\mathbf{C}(\boldsymbol{\chi},\dot{\boldsymbol{\chi}})\|\le \bar{c}\,\|\dot{\boldsymbol{\chi}}\|$, 
$\|\mathbf{g}(\boldsymbol{\chi})\|\le \bar{g}$, 
$\|\mathbf{d}\|\le \bar{d}$.  
\end{property}
The variable dependencies are removed subsequently whenever obvious, e.g., $\mathbf{M}(\boldsymbol{\chi})$ and $\mathbf{C}(\boldsymbol{\chi},\dot{\boldsymbol{\chi}})$ are denoted as M and C for brevity.
\begin{assum} [Uncertainty] \label{assum_1}
The system dynamics terms $\mathbf{M}, \mathbf{C}, \mathbf{g}, \mathbf{d}$ and their bounds $\overline{m}, \underline{m}, \bar{c}, \bar{g}, \bar{d}$ defined in Properties \ref{prop_1}-\ref{prop_2} are unknown for control design. The manipulator Jacobian $\mathbf{J}_\alpha$, which is a kinematic parameter, is assumed to be known.
\end{assum}

The desired trajectories $\boldsymbol{\chi}_d$ and their time-derivatives $\dot{\boldsymbol{\chi}}_d, \ddot{\boldsymbol{\chi}}_d$ are designed to
be bounded. Furthermore, $\boldsymbol{\chi}, \dot{\boldsymbol{\chi}}, \ddot{\boldsymbol{\chi}}$ are considered to be available for feedback along with $\mathbf{F}_{ext}$. Let us define the tracking errors as 
\begin{align}
\mathbf{e} = \boldsymbol{\chi} -\boldsymbol{\chi}_d,
\quad
\dot{\mathbf{e}} = \dot{\boldsymbol{\chi}} - \dot{\boldsymbol{\chi}}_d .\label{err}
\end{align}

\subsection{Control Objective and Challenge}
To ensure compliance under external forces, the aim is to impose a desired impedance behavior across all generalized coordinates via the target impedance dynamics formulated as: 
\begin{equation}\label{Simplified_Impedance_Dynamics}
\boldsymbol{\mathbf{M}}_d\,\ddot{\mathbf{e}}
+
\boldsymbol{\mathbf{K}}_d\,\dot{\mathbf{e}}
+
\boldsymbol{\mathbf{K}}_p\,\mathbf{e}
=
\mathbf{e}_\tau,
\end{equation}
where $\boldsymbol{\mathbf{M}}_d,\boldsymbol{\mathbf{K}}_d,\boldsymbol{\mathbf{K}}_p$ are user-designed constant positive-definite inertia, damping, and stiffness matrices; $\mathbf{e}_\tau = \boldsymbol{\tau}_{\text{ext}} - \boldsymbol{\tau}_d$ represents the net torque deviation from the desired torque $\boldsymbol{\tau}_{d} \triangleq  \begin{bmatrix} \boldsymbol{0}_{3} ~~\boldsymbol{0}_3 ~~\mathbf{J}_\alpha^{\top} \mathbf{F}_{d} \end{bmatrix}^\top \in \mathbb{R}^{6+n} $. 

In practice, due to external disturbances and modeling errors, the actual system does not exactly follow the desired dynamics \eqref{Simplified_Impedance_Dynamics}. The deviation from the desired impedance behavior can be defined as:
\begin{equation}\label{deviation_error} \Delta \mathcal{I} \triangleq \boldsymbol{\mathbf{M}}_d\,\ddot{\mathbf{e}} 
+
\boldsymbol{\mathbf{K}}_d\,\dot{\mathbf{e}}
+
\boldsymbol{\mathbf{K}}_p\,\mathbf{e}
-
\mathbf{e}_\tau.
\end{equation}
where $\Delta \mathcal{I}$, termed as the \emph{impedance error}, captures the discrepancy between the desired impedance model and the actual response.

\begin{remark}[Selection of $\boldsymbol{\mathbf{M}}_d$, $\boldsymbol{\mathbf{K}}_d$ and $\boldsymbol{\mathbf{K}}_p$]
The desired $\boldsymbol{\mathbf{M}}_d$, $\boldsymbol{\mathbf{K}}_d$ and $\boldsymbol{\mathbf{K}}_p$ 
are design parameters selected as per application requirements. $\boldsymbol{\mathbf{M}}_d$ in general is chosen to be a diagonal matrix based on nominal mass and geometric parameters (e.g., link length) of the AAM system, which are often available in practice.
For the states that require compliance, such as those of the manipulator, the corresponding elements in  $\boldsymbol{\mathbf{K}}_d$ and $\boldsymbol{\mathbf{K}}_p$ can be chosen to achieve near-critical damping with lower stiffness, thus enhancing impact absorption. Meanwhile, the elements corresponding to the states of the aerial vehicle, can be chosen to achieve over-damped behaviour with high stiffness. The interested reader is referred to the standard literature for the design of  $\boldsymbol{\mathbf{M}}_d$, $\boldsymbol{\mathbf{K}}_d$ and $\boldsymbol{\mathbf{K}}_p$  \cite{Choosing_impedance_parameters,billard16_impedance}.
\end{remark}

\textbf{Control Problem:} Under Assumption 1 and Properties 1-2, the goal is to design an adaptive impedance controller that compensates for system dynamics uncertainties and disturbances without their a priori knowledge, such that the tracking errors ($\mathbf{e}, \dot{\mathbf{e}}$) and the impedance error $\Delta \mathcal{I}$ remain bounded.

The following section elaborates on the proposed control design and analysis.

\section{Proposed Adaptive Impedance Control and Analysis}
\label{sec:sec3}
\subsection{Impedance Dynamics}
 Let us define an auxiliary error variable $\mathbf{s}$ as
\begin{equation}\label{s_general}
\mathbf{s} = \dot{\mathbf{e}} + \boldsymbol{\Phi}\,\mathbf{e} - \boldsymbol{\gamma},
\end{equation}
where $\boldsymbol{\Phi}$ and $\boldsymbol{\gamma}$ are derived from the following relations:
\begin{subequations} 
\begin{align}
&(\boldsymbol{\mathbf{M}}_d^{-1}\boldsymbol{\mathbf{K}}_d - \boldsymbol{\Phi})~\text{is positive definite,}\label{gain_new}\\
&(\boldsymbol{\mathbf{K}}_d - \boldsymbol{\mathbf{M}}_d \boldsymbol{\Phi}) \boldsymbol{\Phi}  = \boldsymbol{\mathbf{K}}_p , \label{gain_cond}\\
& \dot{\boldsymbol{\gamma}} = \boldsymbol{\mathbf{M}}_d^{-1} (\mathbf{e}_\tau +(\boldsymbol{\mathbf{K}}_d  - \boldsymbol{\mathbf{M}}_d \boldsymbol{\Phi})) \boldsymbol{\gamma} ~~\text{with}~~ \boldsymbol{\gamma}(0) > 0.\label{adaptive_1}
\end{align}
\end{subequations}

The condition (\ref{gain_new}) is used for ensuring closed-loop stability (cf. discussion after (\ref{final})). Multiplying the time derivative of \eqref{s_general} by $\boldsymbol{\mathbf{M}}_d$ and using \eqref{Simplified_Impedance_Dynamics} gives
\begin{equation}\label{step1} \Delta \mathcal{I} = \boldsymbol{\mathbf{M}}_d \dot{\mathbf{s}} + (\boldsymbol{\mathbf{K}}_d - \boldsymbol{\mathbf{M}}_d \boldsymbol{\Phi}) \dot{\mathbf{e}} + \boldsymbol{\mathbf{M}}_d \dot{\boldsymbol{\gamma}} + \boldsymbol{\mathbf{K}}_p \mathbf{e} - \mathbf{e}_\tau
\end{equation}
Substituting $\dot{\mathbf{e}}$ and $\dot{\boldsymbol{\gamma}}$ from \eqref{s_general} and (\ref{adaptive_1}) into \eqref{step1} yields
\begin{align} 
\Delta \mathcal{I} =& \boldsymbol{\mathbf{M}}_d \dot{\mathbf{s}} + (\boldsymbol{\mathbf{K}}_d - \boldsymbol{\mathbf{M}}_d \boldsymbol{\Phi}) \mathbf{s} - (\boldsymbol{\mathbf{K}}_d - \boldsymbol{\mathbf{M}}_d \boldsymbol{\Phi}) \boldsymbol{\Phi} \mathbf{e} + \boldsymbol{\mathbf{K}}_p \mathbf{e} \nonumber \\
-&  (\boldsymbol{\mathbf{K}}_d - \boldsymbol{\mathbf{M}}_d \boldsymbol{\Phi}) \boldsymbol{\gamma} + \boldsymbol{\mathbf{M}}_d  \dot{\boldsymbol{\gamma}} -  \mathbf{e}_\tau\nonumber\\
    =& \boldsymbol{\mathbf{M}}_d \dot{\mathbf{s}} + (\boldsymbol{\mathbf{K}}_d - \boldsymbol{\mathbf{M}}_d \boldsymbol{\Phi}) \mathbf{s}. \label{I_cl}
\end{align} 
%\subsection{Trajectory tracking}
The AAM dynamics \eqref{General_EL_dynamics} can be rearranged as
\begin{equation} \label{uncertain_dynamics}
\boldsymbol{\mathbf{M}}_d\,\ddot{\boldsymbol{\chi}} +\boldsymbol{\mathcal{E}}  = \boldsymbol{\tau} + \boldsymbol{\tau}_{\text{ext}},
\end{equation}
where $\boldsymbol{\mathcal{E}}  \triangleq  (\mathbf{M}-\boldsymbol{\mathbf{M}}_d)\ddot{\boldsymbol{\chi}} +  \mathbf{C}\dot{\boldsymbol{\chi}} + \mathbf{g} + \mathbf{d} $ is treated as the \emph{overall} uncertainty. Multiplying the time derivative of \eqref{s_general} with $\boldsymbol{\mathbf{M}}_d$ and using \eqref{uncertain_dynamics} results into the following:
\begin{align} \label{step3}
 \boldsymbol{\mathbf{M}}_d \dot{\mathbf{s}} = \boldsymbol{\tau} + \boldsymbol{\tau}_{\text{ext}} -  \boldsymbol{\mathcal{E}} - \boldsymbol{\mathbf{M}}_d\Bigl(\ddot{\boldsymbol{\chi}}_{d} - \boldsymbol{\Phi}\,\dot{\mathbf{e}}\Bigr) - \boldsymbol{\mathbf{M}}_d\dot{\boldsymbol{\gamma}}.
\end{align}
Further, using \eqref{adaptive_1} and $\boldsymbol{\tau}_d = \boldsymbol{\tau}_{\text{ext}} - \mathbf{e}_\tau$ in \eqref{step3} gives the closed-loop dynamics
\begin{align} \label{step4}
 \boldsymbol{\mathbf{M}}_d \dot{\mathbf{s}} &=  \boldsymbol{\tau} +  \boldsymbol{\tau}_d  - \boldsymbol{\mathcal{E}}- \boldsymbol{\mathbf{M}}_d(\ddot{\boldsymbol{\chi}}_{d} - \boldsymbol{\Phi}\,\dot{\mathbf{e}}) \nonumber \\
 &+ (\boldsymbol{\mathbf{K}}_d  - \boldsymbol{\mathbf{M}}_d \boldsymbol{\Phi}). 
\end{align}
\subsection{Overall Control Law}
We propose the control input
\begin{subequations} \label{control_law}
\begin{align} 
\boldsymbol{\tau} =&    -\boldsymbol{\Lambda} \mathbf{s} - \boldsymbol{\tau}_d + \Delta\boldsymbol{\tau} +  \boldsymbol{\mathbf{M}}_d(\ddot{\boldsymbol{\chi}}_{d} - \boldsymbol{\Phi}\,\dot{\mathbf{e}}) \nonumber \\
 &- (\boldsymbol{\mathbf{K}}_d  - \boldsymbol{\mathbf{M}}_d \boldsymbol{\Phi}) 
\label{tau} \\
\Delta\boldsymbol{\tau}
&=\;
\begin{cases}
\rho\,\dfrac{\mathbf{s}}{\|\mathbf{s}\|} 
& \text{if } \|\mathbf{s}\|\;\ge\;\varpi,\\[6pt] 
\rho\,\dfrac{\mathbf{s}}{\varpi} 
& \text{if } \|\mathbf{s}\|\;<\;\varpi,
\end{cases} \label{del_tau} 
\end{align}
\end{subequations}
where \(\boldsymbol{\Lambda}\) is a user-defined positive definite gain matrix, $\varpi > 0$ is a scalar used to avoid control chattering, and $\rho$ tackles the overall uncertainties, whose design is now discussed later. 

Using system Properties 1-2, an upper bound structure for $||\mathbf{\mathcal{E}} ||$ can be derived as
\begin{align} \label{boundness}
||\mathbf{\mathcal{E}} ||  \leq \|\mathbf{M}-\boldsymbol{\mathbf{M}}_d\| \|\ddot{\boldsymbol{\chi}}\| + \bar{c} \|\dot{\boldsymbol{\chi}}\|^2 + \bar{g} + \bar{d}.
\end{align}
Defining $\boldsymbol{\xi} \triangleq [\mathbf{e}^\top~ \dot{\mathbf{e}}^\top]^\top$ , using the inequalities $||\boldsymbol{\xi}|| \geq ||\mathbf{e}||$ and $||\boldsymbol{\xi}|| \geq ||\dot{\mathbf{e}}||$, and substituting $\dot{\boldsymbol{\chi}} = \dot{\mathbf{e}} + \dot{\mathbf{\chi}}_d$ in (\ref{boundness}) yields
\begin{align}
 ||\mathbf{\mathcal{E}} ||  &\leq   \mathcal{H}_{0}^*  + \mathcal{H}_{1}^*\|\boldsymbol{\xi}\| + \mathcal{H}_{2}^*\|\boldsymbol{\xi}\|^2 + \mathcal{H}_{3}^*\|\ddot{\boldsymbol{\chi}}\|\label{up_bound_E}  
\\
\text{where} \quad \mathcal{H}_{0}^* &= 
(\bar{g} + \bar{d} + \bar{c} ||\dot{\boldsymbol{\chi}}_{d} ||^2); ~~\mathcal{H}_{1}^* = 2 \bar{c} || \dot{\boldsymbol{\chi}}_{d} || \nonumber \\
\mathcal{H}_{2}^* &=  \bar{c};  ~~\mathcal{H}_{3}^* =  ||\mathbf{M}-\boldsymbol{\mathbf{M}}_d || \nonumber
\end{align}
are \emph{unknown scalars} due to system uncertainties. Based on the upper bound structure in (\ref{up_bound_E}), the gain $\rho$ in (\ref{del_tau}) is thus designed as
\begin{align}
\rho &= \hat{\mathcal{H}}_{0} + \hat{\mathcal{H}}_{1}||\boldsymbol{\xi}|| + \hat{\mathcal{H}}_{2}||\boldsymbol{\xi}||^2 + \hat{\mathcal{H}}_{3}||\ddot{\boldsymbol{\chi}}|| + \zeta 
\end{align}
where $\zeta$ is an auxiliary stabilizing gain and $\hat{\mathcal{H}}_i$ are the estimates of ${\mathcal{H}}_i^*$ for each $i=0,1,2,3$. These gains are adapted via the following laws:
\begin{subequations} \label{adaptive_law_full}
\begin{align}
&\dot{\hat{\mathcal{H}}}_{i} = ||\mathbf{s}|| ||\boldsymbol{\xi}||^i - \nu_{i} \hat{\mathcal{H}}_{i},~~ i = 0,1,2, \\
&\dot{\hat{\mathcal{H}}}_{3} = \|\mathbf{s}\|\|\ddot{\boldsymbol{\chi}}\| - \nu_{3} \hat{\mathcal{H}}_{3}, \\
&\dot{\zeta} = \begin{cases}
    0, ~~~~~~~~~ \text{if } \|\mathbf{s}\| \geq \varpi,  \\
    -\left( 1 + \left(\hat{\mathcal{H}}_{3} \|\ddot{\boldsymbol{\chi}}\|+ \sum_{i=0}^{2} \hat{\mathcal{H}}_{i} \|\boldsymbol{\xi}\|^i \right) \|\mathbf{s}\| \right) \zeta + {\epsilon}, \label{zeta}  \\
    ~~~~~~~~~~~~~ \text{if } \|\mathbf{s}\| < \varpi,
\end{cases} \\
& \qquad ~\hat{\mathcal{H}}_{i} (0) > 0,~\hat{\mathcal{H}}_{3} (0) > 0,~\zeta (0) > 0, 
\end{align}
\end{subequations}
where $\nu_{ 0}, \nu_{ 1}, \nu_{ 2}, \nu_{ 3}, {\epsilon} \in\mathbb{R}^{+}$ are user-defined scalars. The closed-loop stability result is stated below.

\begin{theorem}
Under Assumption \ref{assum_1} and Properties~\ref{prop_1}--\ref{prop_2}, the closed-loop trajectories of \eqref{step4} with control law \eqref{control_law} with adaptive law \eqref{adaptive_1} and \eqref{adaptive_law_full} remains uniformly ultimately bounded (UUB), implying boundedness of $\mathbf{e}, \dot{\mathbf{e}}, \Delta \mathcal{I}$.
\end{theorem}
\textit{Proof:}
 See Appendix.

%%%%%%%%%%%%%%%%%%%%%%%%%%%%%%%%%%%%%%%%%%%%%%%%%%%%%%%%%%%%%%%%%%%%%%%%%%%%%%%%
\section{Experimental Results and Analysis}
\label{sec:sec4}

\subsection{Experimental Setup}
Experiments were performed on a AAM comprising a Tarot Ironman 650 quadcopter (SunnySky V4006 motors) and a 2R serial-link manipulator (Dynamixel XM430-W210-T motors). A net with a 6-axis force-torque sensor (RFT60-HA01) is attached to the manipulator for active feedback during payload catching. The overall system weighs approximately 3.0 kg, powered via a U2D2 Power Hub and controlled by a Raspberry Pi-4 interfaced through a U2D2 converter. Optitrack motion capture system and IMU data are fused to obtain the system states needed for feedback. Trajectory setpoints for the quadrotor and manipulator are pre-computed using \textit{mav\_trajectory\_generation}~\cite{richter2016polynomial} and \textit{ruckig}~\cite{berscheid2021jerk}, respectively.

\begin{figure}[!h]
\centering
\includegraphics[width=0.45\textwidth]{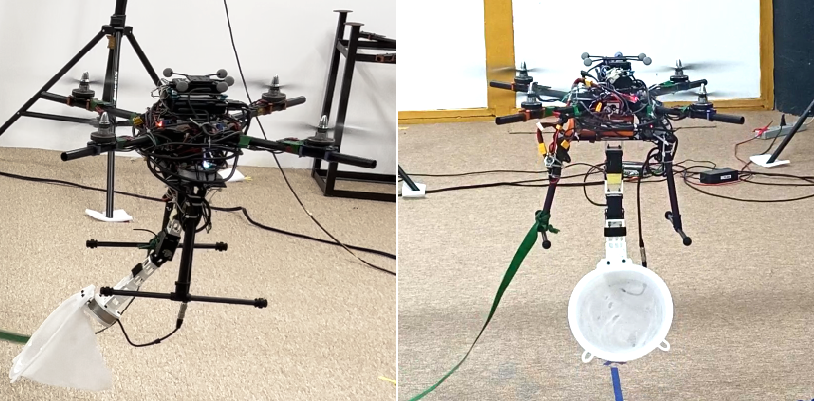}
\caption{Hardware setup for the AAM.}
\label{comparison}
\end{figure}

%%%%%%%%%%%%%%%%%%%

\begin{table}[htbp]
\centering
\caption{\textcolor{black}{Design Parameters for the Proposed Controller}}
\renewcommand{\arraystretch}{1.3} 
\resizebox{\columnwidth}{!}{%
\begin{tabular}{|l|l|}
\hline
\textbf{Parameter} & \textbf{Value} \\ \hline
\({\mathbf{M}}_d\)& \(\mathrm{diag}\{1,\,1,\,1,\;0.015,\,0.015,\,0.015,\;0.1,\,0.1\}\) \\ \hline
\(\boldsymbol{\Lambda}\) & \(\mathrm{diag}\{1.5,\,1.5,\,2.0,\;3.5,\,3.5,\,2.5,\;1.0,\,1.0\}\) \\ \hline
\(\boldsymbol{\Phi}\) & \(\mathrm{diag}\{4.0,\,4.0,\,4.0,\;3.0,\,3.0,\,3.0,\;3.5,\,3.5\}\) \\ \hline
\(\boldsymbol{\mathbf{K}}_d\)& \(\mathrm{diag}\{40,\,40,\,40,\;0.75,\,0.75,\,0.75,\;0.7,\,0.7\}\)\\ \hline
\(\boldsymbol{\mathbf{K}}_p\)& \(\mathrm{diag}\{144,\,144,\,144,\;2.115,\,2.115,\,2.115,\;1.225,\,1.225\}\)\\ \hline
Initial adaptive gains & \(\hat{\mathcal{H}}_0(0)=\; \hat{\mathcal{H}}_1(0)=\; \hat{\mathcal{H}}_2(0)=\; \hat{\mathcal{H}}_3(0)=0.01\)\\ \hline
Initial auxiliary gain & \(\zeta(0)=0.1\) \\ \hline
Adaptation rates & \(\nu_0=\nu_1=\nu_2=\nu_3=10.0\) \\ \hline
\(\epsilon\) & \(0.0001\) \\ \hline
Boundary layer & \(\varpi = 0.1\) \\ \hline
\end{tabular}%
}
\label{tab:control_parameters}
\end{table}

\subsection{Experimental Scenarios, Results, and Analysis}\label{sec:experiments}
\subsubsection{{Scenario}}
The AAM is tasked with catching payloads of various mass (in the range $0.1$ - $0.3$ kg) while following a straight-line trajectory. The experimental scenario is described below:
\begin{figure*}[t]
\centering
\includegraphics[width=\textwidth]{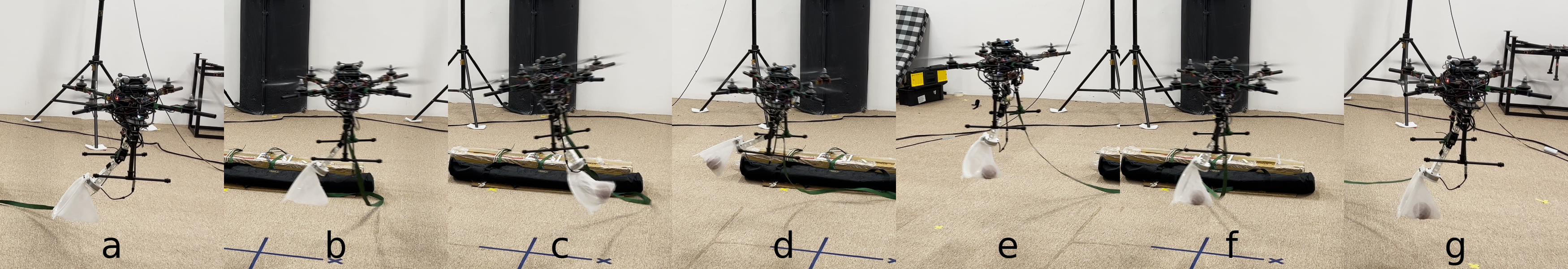}
\caption{Sequence of operations with the proposed controller: 
(a)~Takeoff,
(b)~Follows trajectory while arm expands to pick payload,
(c)~Payload catching during flight and stabilization,
(d)~Transporting the payload,
(e)-(g)~Return to original position.  
The quadrotor is tethered for safety.}
\label{fig:collage}
\end{figure*}

% \vspace{2pt}

\begin{itemize}
\item  The quadrotor takes off from $(x,y) = (-1,0)$ with the manipulator links at angles $(\alpha_1,\alpha_2) = (\ang{165}, \ang{140})$, and ascends to  $z=1$ m.
\item A straight-line trajectory to  $(x,y)=(1,0)$ is commanded, followed by return to the trajectory start point, maintaining $z=1$ m.  
\item While following the straight line, a payload is thrown into the net while the AAM is passing through $(x,y)=(0,0)$ at approx. $t = 6.5$ sec.  

\item After catching the payload, the AAM completes the remaining portion of the trajectory and returns to its starting position.
\end{itemize}

\textbf{Performance comparisons:} To suitably judge the importance and performance of the proposed controller, we have selected a Complete System Compliance (CSC) based method as in \cite{Cataldi_16} and a Partitioned System Compliance (PSC) based method as in \cite{wang_impact_2024} for comparison. The method \cite{Cataldi_16} aims to achieve complete system compliant behavior; whereas, \cite{wang_impact_2024} only aims for arm compliance with the quadrotor being controlled ignoring the coupling forces. Both these controllers does not take into account system parametric uncertainties; however \cite{wang_impact_2024} estimates the payload mass. Therefore, these state-of-the-art methods act as suitable platform to verify the importance of the proposed method in providing complete system compliance under parametric uncertainty. 

The performance of these three controllers are evaluated under the aforementioned scenario for three different payloads of $0.1$ kg, $0.2$ kg and $0.3$ kg and the payload mass information kept unknown for all. The control parameters for \cite{Cataldi_16} and \cite{wang_impact_2024} are taken from the respective works after adjusting for the AAM parameters. Note that the proposed method does not require any knowledge of system parameters (including the payload mass) for control design, and the same control parameters are selected for all three payloads (listed in Table~\ref{tab:control_parameters}).

\subsubsection{Results and Analysis}
\begin{figure}[t]
\centering
\includegraphics[width=0.5\textwidth]{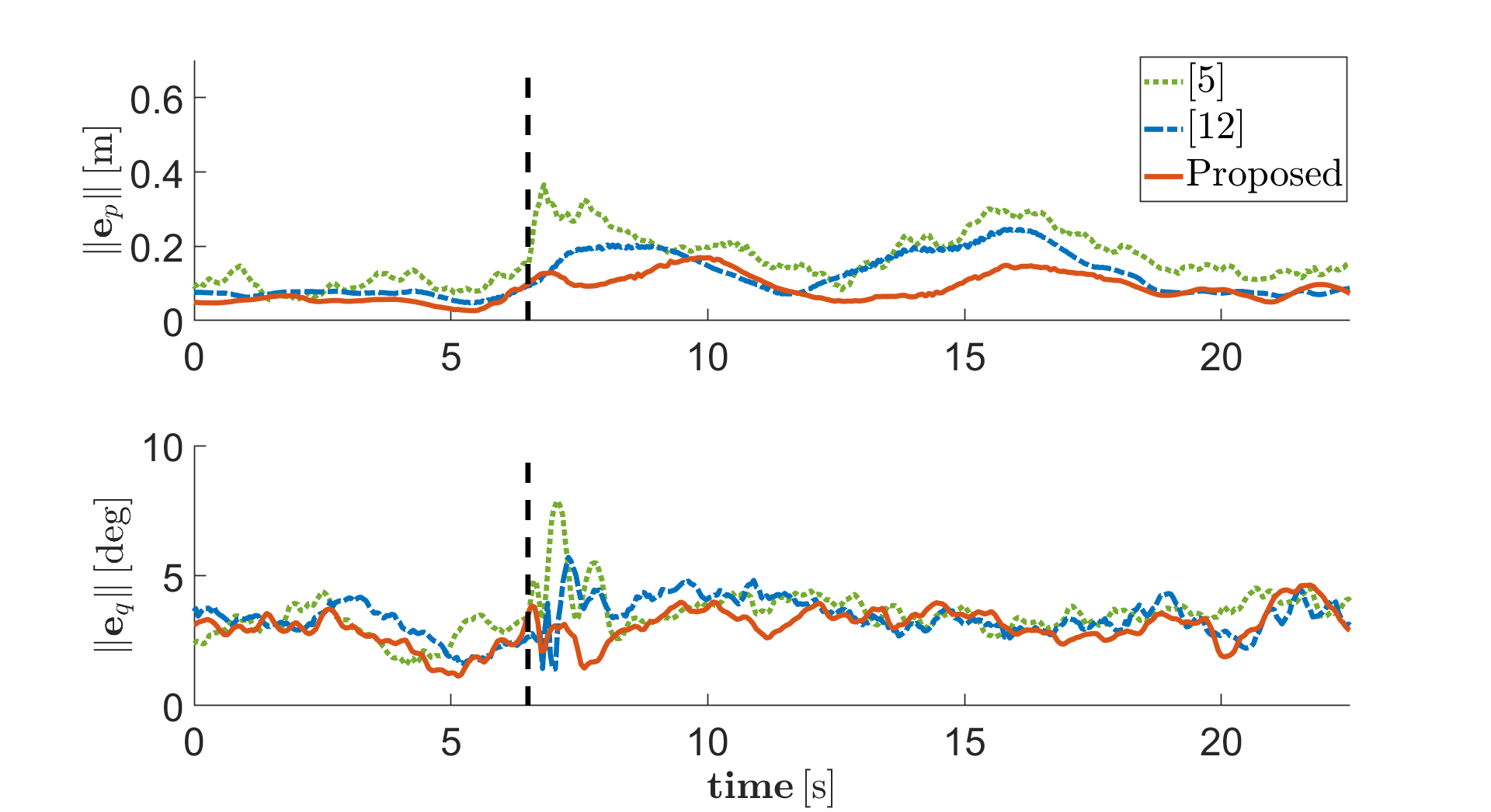}
\caption{Tracking error comparison for $0.1$ kg payload (the vertical black like denotes the time the payload is caught).}
\label{pq_100g}
\end{figure}

\begin{figure}[t]
\centering
\includegraphics[width=0.5\textwidth]{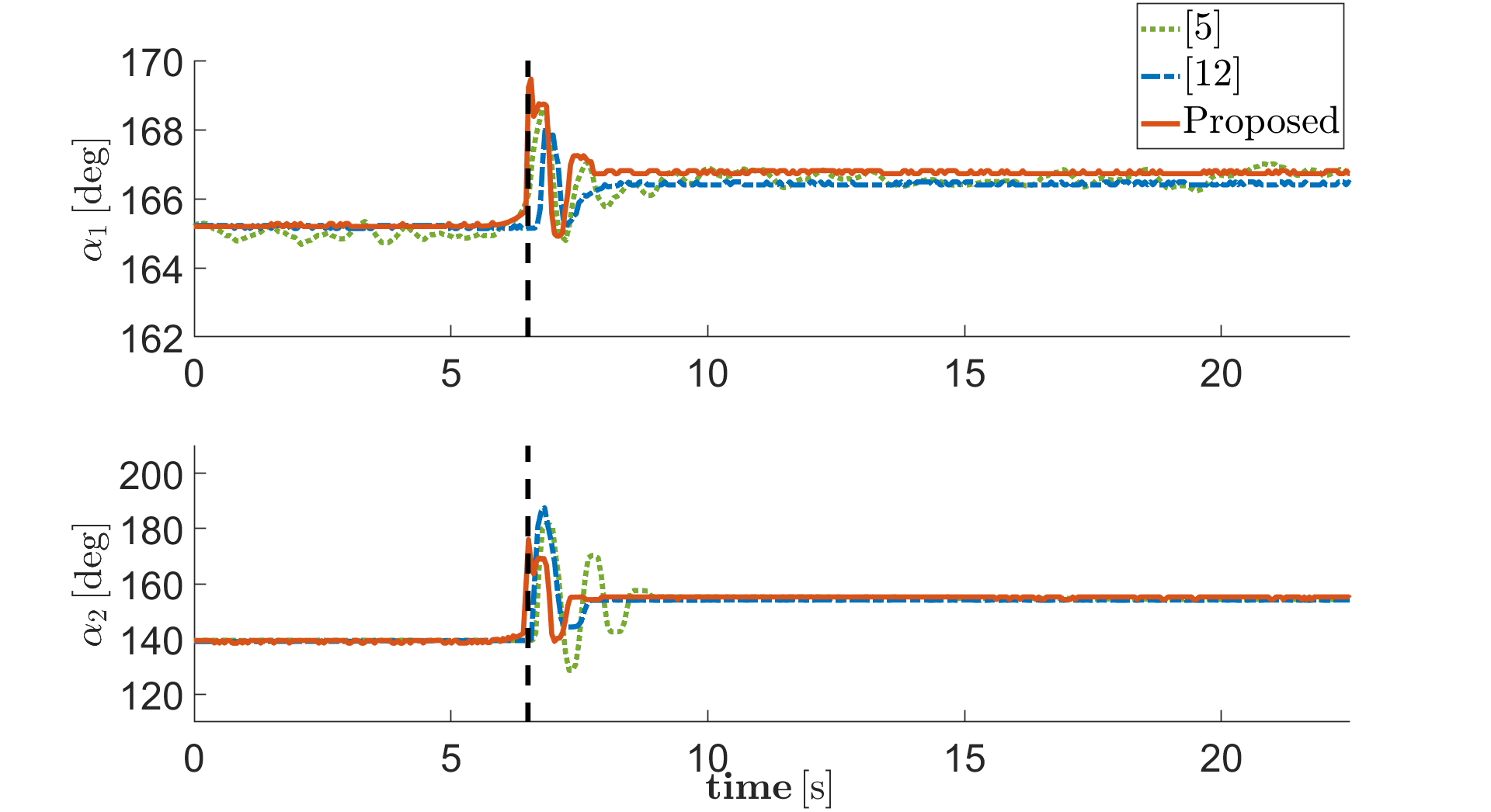}
\caption{Arm angles for $0.1$ kg payload (the vertical black like denotes the time the payload is caught).}
\label{motor_100g}
\end{figure}

\begin{figure}[t]
\centering
\includegraphics[width=0.5\textwidth]{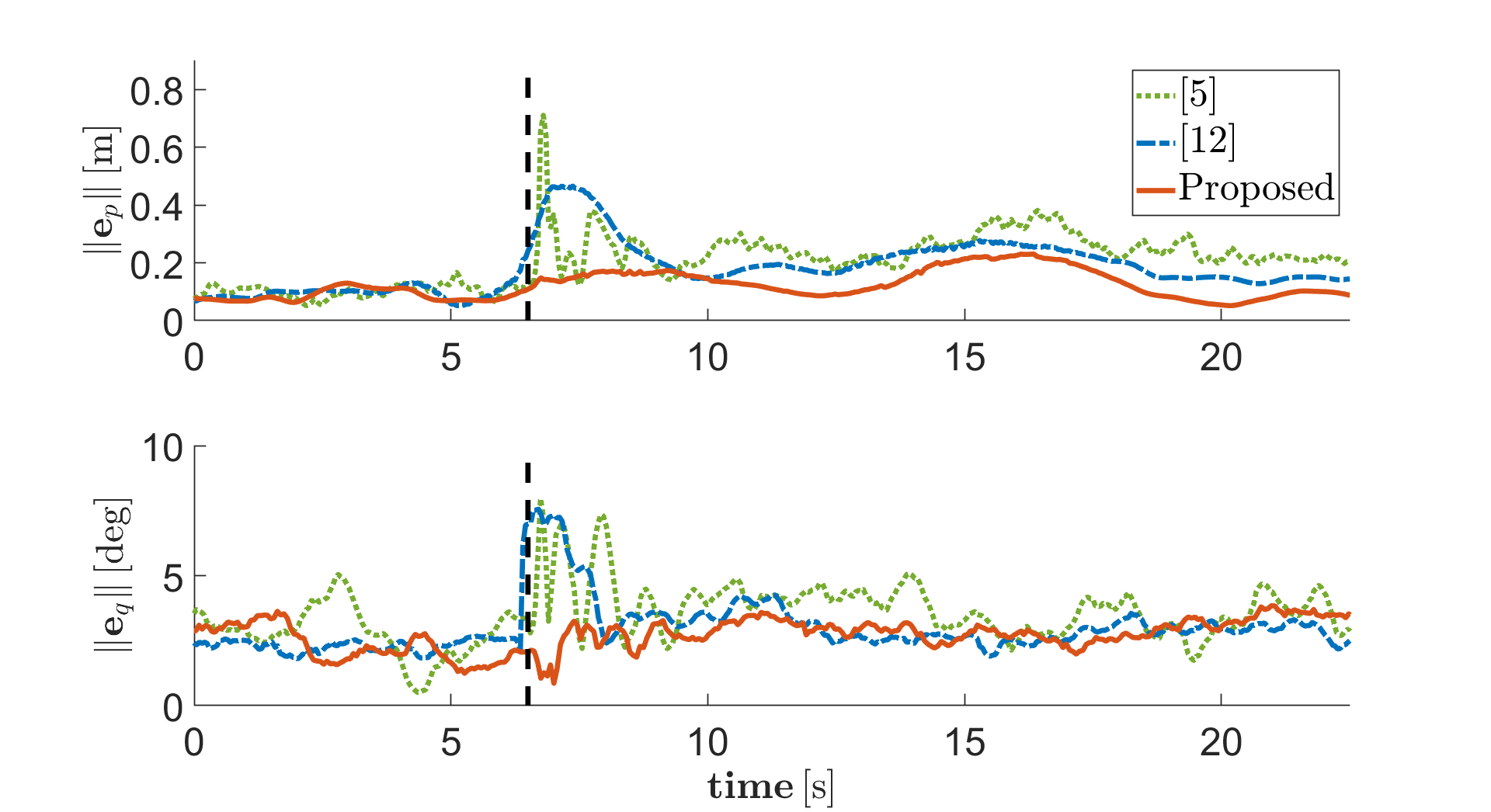}
\caption{Tracking error comparison for $0.2$ kg payload (the vertical black like denotes the time the payload is caught).}
\label{pq_200g}
\end{figure}

\begin{figure}[t]
\centering
\includegraphics[width=0.5\textwidth]{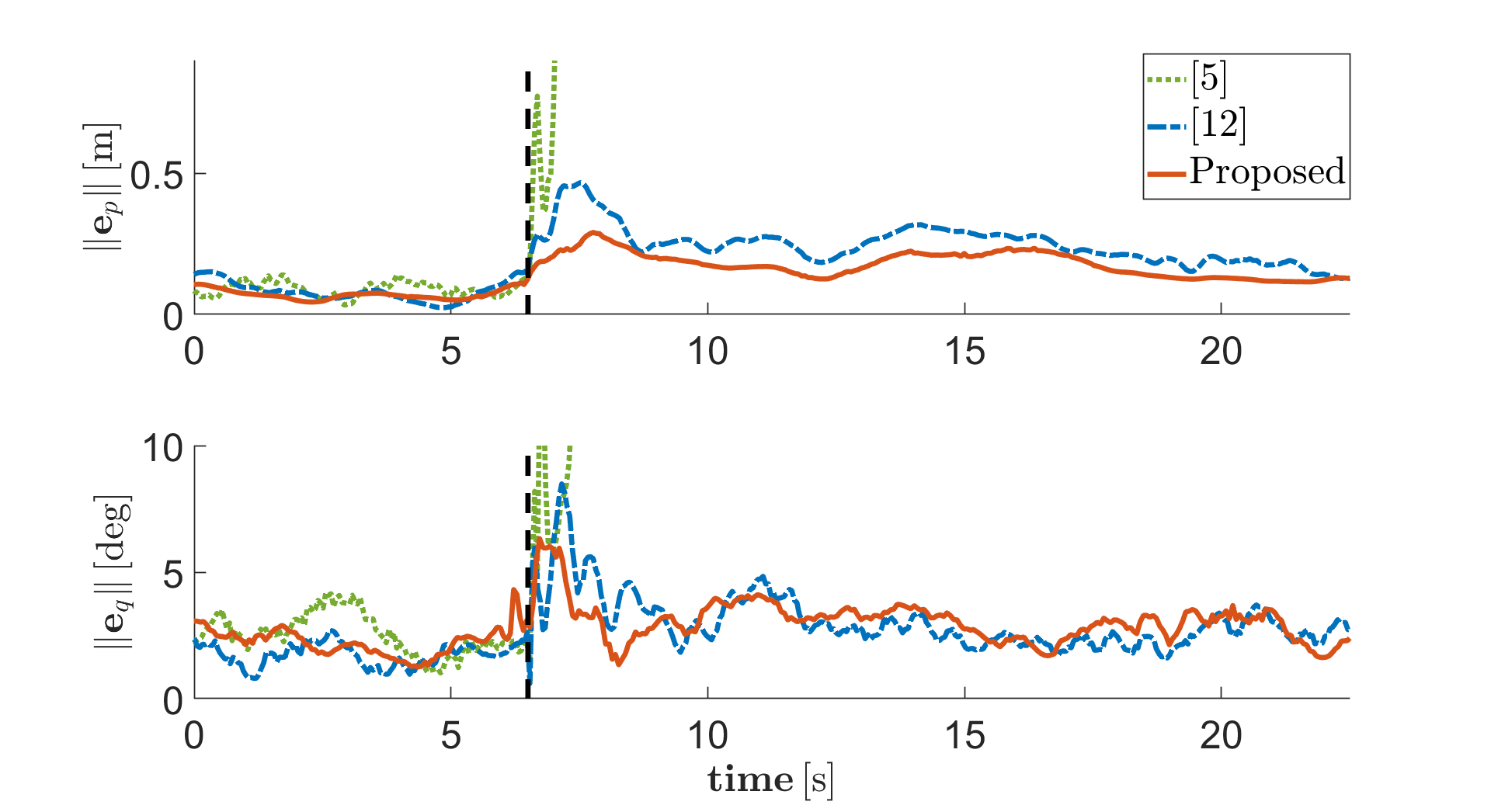}
\caption{Tracking error comparison for $0.3$ kg payload (the vertical black like denotes the time the payload is caught).}
\label{pq_300g}
\end{figure}

\begin{figure}[t]
\centering
\includegraphics[width=0.5\textwidth]{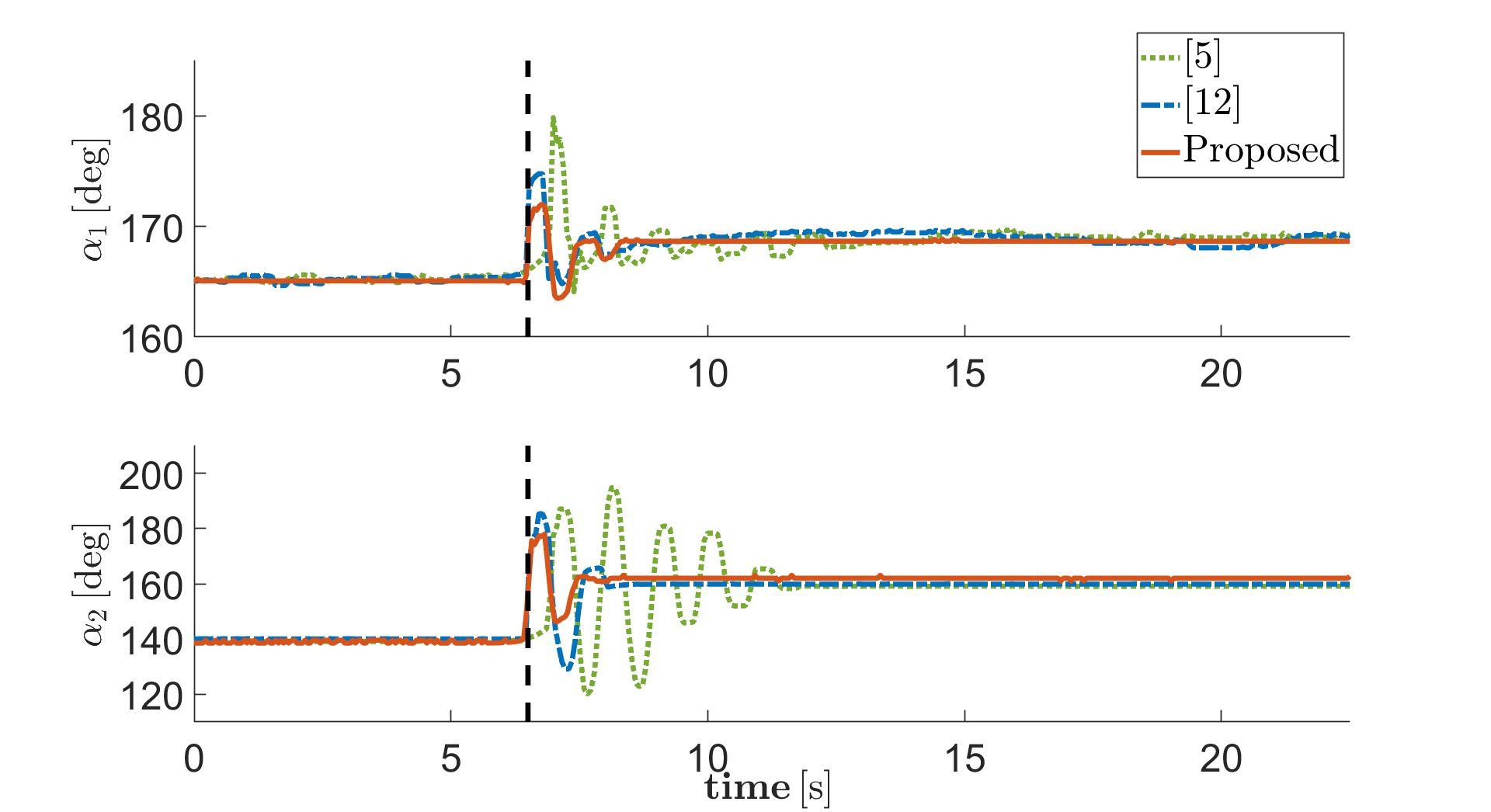}
\caption{Arm angles for $0.2$ kg payload (the vertical black like denotes the time the payload is caught).}
\label{motor_200g}
\end{figure}

\begin{figure}[t]
\centering
\includegraphics[width=0.5\textwidth]{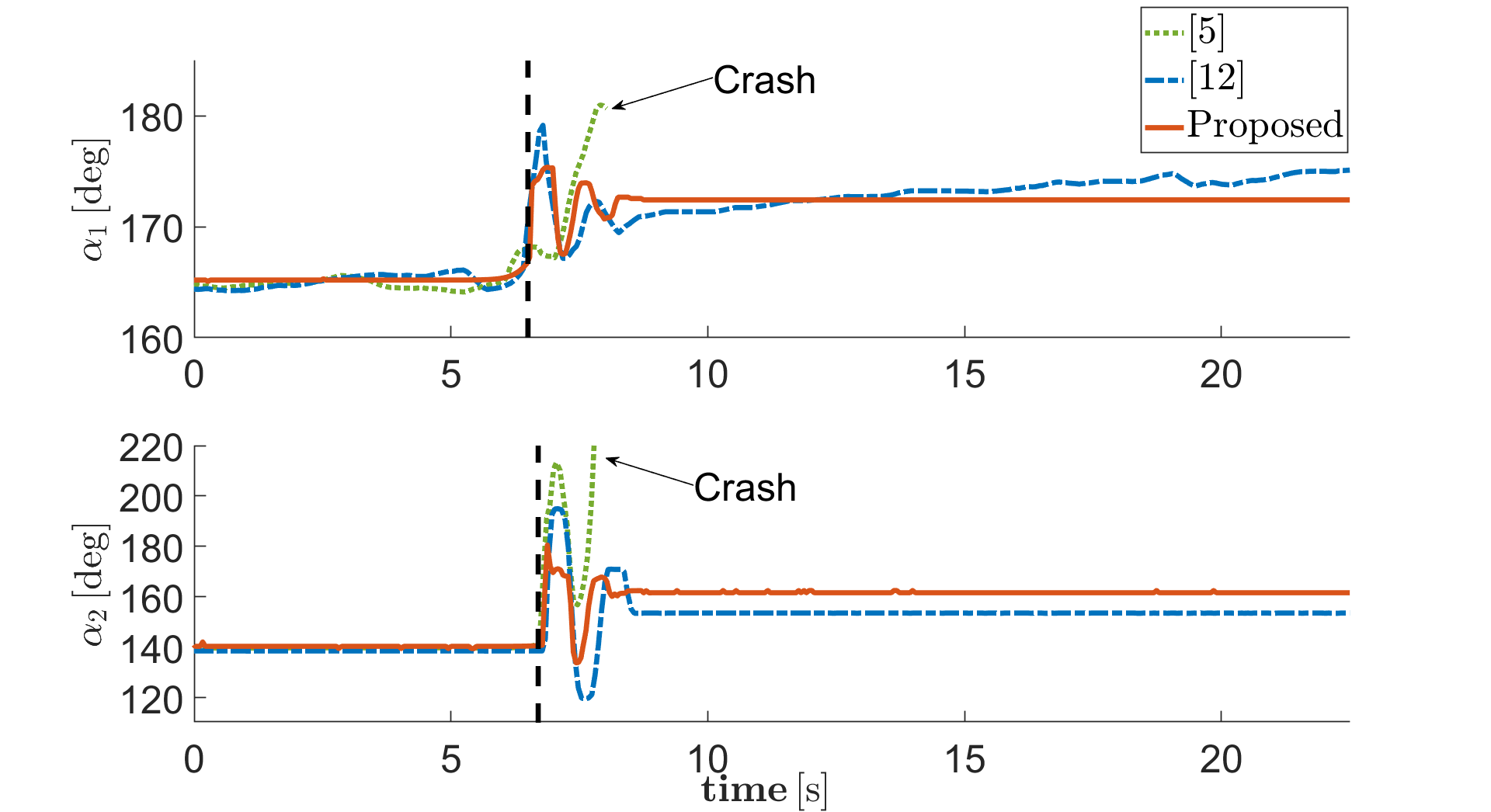}
\caption{Arm angles for $0.3$ kg payload. Incomplete green lines indicate the crash with \cite{Cataldi_16} (the vertical black like denotes the time the payload is caught).}
\label{motor_300g}

\end{figure}

The tracking performance of the three controllers are provided in Figs.~\ref{pq_100g}-\ref{motor_300g} in terms of position error norm $\|e_p\|$, orientation error norm $\|e_q\|$ and manipulator arm angles for three different payloads. The vertical lines in the figures denote the time of catching the payload. Further, for better inference of the quadrotor behavior, Table \ref{RMS_table} collects the root mean squared (RMS) values of $\|e_p\|$ and $\|e_q\|$ pre-catching and post-catching.

The error plots reveal that the quadrotor pose shows oscillatory behavior for all controllers around the time of impact with the payload, and such deviations increase as the payload mass increases. In fact, the AAM with \cite{Cataldi_16} crashes after the impact with $0.3$ kg payload (cf. the broken green lines in Figs. \ref{pq_300g} and \ref{motor_300g}); this happens because with increased payload the inertial forces, especially the coupling forces, become more pronounced. Since \cite{Cataldi_16} is not equipped with any uncertainty handling mechanism, its performance suffers. Having a payload mass estimation method, \cite{wang_impact_2024} performs better than \cite{Cataldi_16}. Nevertheless, ignorance of coupling forces between the quadrotor and the arm significantly deteriorates its performance compared to the proposed controller; this is clearly evident in the quadrotor position deviation for \cite{wang_impact_2024} in Fig. \ref{pq_300g}. Overall, the proposed controller with its adaptive nature to tackle uncertainties without their a priori knowledge significantly outperforms the state of the art \cite{Cataldi_16} and \cite{wang_impact_2024}.

\begin{table*}[t]
\centering
\caption{Tracking Performance Comparison in terms of RMS errors}
\label{RMS_table}
\resizebox{\linewidth}{!}{
\begin{tabular}{c|cccc|cccc|cccc}
\toprule
 & \multicolumn{4}{c|}{CSC Controller {\cite{Cataldi_16}}} & \multicolumn{4}{c|}{PSC Controller {\cite{wang_impact_2024}}} & \multicolumn{4}{c}{Proposed adaptive controller}\\
Object & \multicolumn{2}{c}{Pre-catching} & \multicolumn{2}{c|}{Post-catching} & \multicolumn{2}{c}{Pre-catching} & \multicolumn{2}{c|}{Post-catching} & \multicolumn{2}{c}{Pre-catching} & \multicolumn{2}{c}{Post-catching}\\
\cmidrule(lr){2-5} \cmidrule(lr){6-9} \cmidrule(lr){10-13}
 & $\|e_p\|$ & $\|e_q\|$ & $\|e_p\|$ & $\|e_q\|$ & $\|e_p\|$ & $\|e_q\|$ & $\|e_p\|$ & $\|e_q\|$ & $\|e_p\|$ & $\|e_q\|$ & $\|e_p\|$ & $\|e_q\|$\\
\midrule
0.1 kg & 0.0998& 3.2985& 0.2300& 3.7644& 0.0837& 3.2516& 0.1497& 3.1340& 0.0858& 2.8543& 0.0984& 3.0710\\[0.8ex]
0.2 kg & 0.0989 & 3.0888 & 0.2927 & 3.8446 & 0.0928 & 2.5393 & 0.2236 & 3.2612 & 0.0906 & 2.6097 & 0.1654 & 3.1190\\[0.8ex]
0.3 kg & 0.0932 & 2.8178 & \multicolumn{2}{c|}{Crashed}& 0.0990& 2.4189& 0.2865& 3.5753& 0.0726 & 2.6866& 0.1728 & 3.2031\\
\bottomrule
\end{tabular}
}
\end{table*}

%%%%%%%%%%%%%%%%%%%%%%%%%%%%%%%%%%%%%%%%%%%%%%%%%%%%%%%%%%%%%%%%%%%%%%%%%%%%%%%%
\section{Conclusion}
\label{sec:sec5}

An adaptive impedance controller for complete AAM control was introduced and tested for an object catching scenario. The closed-loop stability of the proposed controller for the AAM was established using the Lyapunov theory. The proposed controller, in contrast to the state-of-the-art, was designed to establish a compliant behavior while tackling unknown state-dependent uncertainty without their prior knowledge. The proposed design was noted to significantly outperform the state-of-the-art approaches for the chosen experimental scenario.
Future work includes incorporating real-time optimization of impedance parameters to extend this adaptive architecture to tasks like collaborative payload transportation and dynamic contact with rigid surfaces.

%%%%%%%%%%%%%%%%%%%%%%%%%%%%%%%%%%%%%%%%%%%%%%%%%%%%%%%%%%%%%%%%%%%%%%%%%%%%%%%%

\section*{APPENDIX: PROOF OF THEOREM 1}
\label{sec:appendix}
Considering the dynamical laws \eqref{adaptive_1} and \eqref{adaptive_law_full} as first-order linear time-varying systems, and using their analytical solutions from positive initial conditions, it can be verified
that $\boldsymbol{\gamma} \geq 0$ $\forall t>0$ and $\hat{\mathcal{H}}_i\geq 0$ (for $i=0,\cdots,3$)  and $\exists \bar{\zeta},  \underline{\zeta} \in \mathbb{R}^{+}$ such that
\begin{align} \label{bound_zeta}
0 < \underline{\zeta} \leq \zeta (t) \leq \bar{\zeta}, ~~ \forall t > 0.     
\end{align}
Stability is analyzed using the following Lyapunov function:

\begin{equation}\label{lyp}
\mathcal{V} = \frac{1}{2}\,\mathbf{s}^\top \boldsymbol{\mathbf{M}}_d \mathbf{s} + \sum_{i=0}^{3}\frac{1}{2}\Bigl(\hat{\mathcal{H}}_i - \mathcal{H}_i^*\Bigr)^2 + \frac{\zeta}{\underline{\zeta}},
\end{equation}
Using (\ref{step4}) and (\ref{tau}), the time derivative of (\ref{lyp}) yields

\begin{align} \label{V_dot}
\dot{\mathcal{V}} &= \mathbf{s}^\top \boldsymbol{\mathbf{M}}_d\,\dot{\mathbf{s}} + \sum_{i=0}^{3} \Bigl(\hat{\mathcal{H}}_i - \mathcal{H}_i^*\Bigr) \dot{\hat{\mathcal{H}}}_i + \frac{\dot{\zeta}}{\underline{\zeta}} \nonumber \\
&= \mathbf{s}^\top \Bigl( -\boldsymbol{\Lambda} \mathbf{s}  - \boldsymbol{\mathcal{E}} + \Delta\boldsymbol{\tau} \Bigr) + \sum_{i=0}^{3} \Bigl(\hat{\mathcal{H}}_i - \mathcal{H}_i^*\Bigr) \dot{\hat{\mathcal{H}}}_i + \frac{\dot{\zeta}}{\underline{\zeta}}
\end{align}
The time derivative (\ref{V_dot}) is now analyzed along with two cases in (\ref{del_tau}).

\textbf{Case (i): \(\|\mathbf{s}\| \ge \varpi\).}\\
From \eqref{V_dot} yields
\begin{align} \label{lyap_2}
\dot{\mathcal{V}} \leq& - \lambda_{\min}(\boldsymbol{\Lambda})||\mathbf{s}||^2  + ||\boldsymbol{\mathcal{E}}|| ||\mathbf{s}|| - \rho ||\mathbf{s}|| 
\nonumber \\
+& \sum_{i=0}^{3} \Bigl(\hat{\mathcal{H}}_i - \mathcal{H}_i^*\Bigr) \dot{\hat{\mathcal{H}}}_i
\nonumber \\
\leq& - \lambda_{\min}(\boldsymbol{\Lambda})||\mathbf{s}||^2 -\sum_{i=0}^{2}  \Bigl(\hat{\mathcal{H}}_i - \mathcal{H}_i^*\Bigr) ||\boldsymbol{\xi}||^i ||\mathbf{s}|| \nonumber \\ 
-& \Bigl(\hat{\mathcal{H}}_3 - \mathcal{H}_3^*\Bigr)  ||\ddot{\boldsymbol{\chi}}|| ||\mathbf{s}||  + \sum_{i=0}^{3} \Bigl(\hat{\mathcal{H}}_i - \mathcal{H}_i^*\Bigr) \dot{\hat{\mathcal{H}}}_i
\end{align}
From (\ref{adaptive_law_full}) we have
 \begin{align} \label{adaptive}
 &\sum_{i=0}^{3} \Bigl(\hat{\mathcal{H}}_i - \mathcal{H}_i^*\Bigr) \dot{\hat{\mathcal{H}}}_i = \sum_{i=0}^{2} \Bigl(\hat{\mathcal{H}}_i - \mathcal{H}_i^*\Bigr) \Bigl( ||\mathbf{s}|| ||\boldsymbol{\xi}||^i - \nu_{i} \hat{\mathcal{H}}_{i}  \Bigr)
 \nonumber \\ 
 & + 
 \Bigl(\hat{\mathcal{H}}_3 - \mathcal{H}_3^*\Bigr) \Bigl( \|\mathbf{s}\|\|\ddot{\boldsymbol{\chi}}\| - \nu_{3} \hat{\mathcal{H}}_{3} \Bigr)
 \nonumber \\ 
 &= \sum_{i=0}^{2}  \Bigl(\hat{\mathcal{H}}_i - \mathcal{H}_i^*\Bigr) ||\boldsymbol{\xi}||^i ||\mathbf{s}|| + \Bigl(\hat{\mathcal{H}}_3 - \mathcal{H}_3^*\Bigr)  ||\ddot{\boldsymbol{\chi}}|| ||\mathbf{s}||  
 \nonumber \\ & + 
 \sum_{i=0}^{3}  \Bigl( \nu_i \hat{\mathcal{H}}_i \mathcal{H}_i^{*} - \nu_i \hat{\mathcal{H}}_i^2\Bigr).  
 \end{align}
 One can verify that
 \begin{equation} \label{ineq}
 \Bigl( \nu_i \hat{\mathcal{H}}_i \mathcal{H}_i^{*} - \nu_i \hat{\mathcal{H}}_i^2 \Bigr) \leq - \frac{\nu_i}{2}  \left((\hat{\mathcal{H}}_i - \mathcal{H}_i^*)^2 - {\mathcal{H}_i^*}^2\right). 
 \end{equation}
Substituting \eqref{adaptive} and \eqref{ineq} into \eqref{lyap_2} yields
\begin{align} \label{lyap_3}
 \dot{\mathcal{V}} \leq& - \lambda_{\min}(\boldsymbol{\Lambda})||\mathbf{s}||^2 - \sum_{i=0}^{3}  \frac{\nu_i}{2}  \left((\hat{\mathcal{H}}_i - \mathcal{H}_i^*)^2 - {\mathcal{H}_i^*}^2\right)
\end{align}
The definition of $\mathcal{V}$ as in \eqref{lyp} yields
\begin{align}  \label{V_bound}
{\mathcal{V}} \leq& - \lambda_{\max}(\boldsymbol{\Lambda})||\mathbf{s}||^2  + \sum_{i=0}^{3}\frac{(\hat{\mathcal{H}}_i - \mathcal{H}_i^*)^2}{2} +\frac{\bar{\zeta}}{\underline{\zeta}}   
\end{align}
Using (\ref{V_bound}), the condition (\ref{lyap_3}) is further simplified to
 \begin{align} \label{lyap_case1}
\dot{\mathcal{V}} \leq -\varrho \mathcal{V} + \varrho \frac{\bar{\zeta}}{\underline{\zeta}}  +  \frac{1}{2}\sum \limits_{i=0}^{3} \nu_{i} {\mathcal{H}^*_{i}}^2
 \end{align}
 where $\varrho \triangleq  \frac{\min( \lambda_{\min}(\boldsymbol{\Lambda}),{\nu_i}/2 )}{\max(\lambda_{\max}(\boldsymbol{\mathbf{M}}_d) ,1/2)} >0$.

\textbf{Case (ii): \(\|\mathbf{s}\| < \varpi\).}\\
In this scenario, from \eqref{V_dot} we have
\begin{align} 
\dot{\mathcal{V}} \leq& - \lambda_{\min}(\boldsymbol{\Lambda})||\mathbf{s}||^2  + ||\boldsymbol{\mathcal{E}}|| ||\mathbf{s}|| - \rho \frac{||\mathbf{s}||^2}{\varpi} 
\nonumber \\
+& \sum_{i=0}^{3} \Bigl(\hat{\mathcal{H}}_i - \mathcal{H}_i^*\Bigr) \dot{\hat{\mathcal{H}}}_i +  \frac{\dot{\zeta}}{\underline{\zeta}}
\nonumber \\
\leq& - \lambda_{\min}(\boldsymbol{\Lambda})||\mathbf{s}||^2  + ||\boldsymbol{\mathcal{E}}|| ||\mathbf{s}|| +  \sum_{i=0}^{3} \Bigl(\hat{\mathcal{H}}_i - \mathcal{H}_i^*\Bigr) \dot{\hat{\mathcal{H}}}_i + \frac{\dot{\zeta}}{\underline{\zeta}}
\nonumber \\
\leq& - \lambda_{\min}(\boldsymbol{\Lambda})||\mathbf{s}||^2  + \sum_{i=0}^{2}\hat{\mathcal{H}}_i |||\boldsymbol{\xi}||^i ||\mathbf{s}|| + \hat{\mathcal{H}}_3 ||\ddot{\boldsymbol{\chi}}|| ||\mathbf{s}|| \nonumber\\
&-\sum \limits_{i=0}^{2} \frac{\nu_i}{2} \left((\hat{\mathcal{H}}_i -  \mathcal{H}_i^*)^2 - {\mathcal{H}_i^*}^2\right)  + \frac{\dot{\zeta}}{\underline{\zeta}}. \label{lyap_4}
\end{align}
The adaptive law (\ref{zeta}) and relation (\ref{bound_zeta}) lead to
 \begin{align} 
 \label{zeta_rep}
\frac{\dot{\zeta}}{\underline{\zeta}} &= -\left( 1 + \left(\hat{\mathcal{H}}_{3} \|\ddot{\boldsymbol{\chi}}\|+ \sum_{i=0}^{2} \hat{\mathcal{H}}_{i} \|\boldsymbol{\xi}\|^i \right) \|\mathbf{s}\| \right)  \frac{{\zeta}}{\underline{\zeta}} + \frac{{{\epsilon}}}{\underline{\zeta}}   \nonumber \\
& \leq   - \hat{\mathcal{H}}_3 ||\ddot{\boldsymbol{\chi}}|| ||\mathbf{s}|| - \sum_{i=0}^{2}\hat{\mathcal{H}}_i |||\boldsymbol{\xi}||^i ||\mathbf{s}|| + \frac{{{\epsilon}}}{\underline{\zeta}}.
 \end{align}
Substituting (\ref{lyap_4}) into (\ref{zeta_rep}) and using (\ref{V_bound}), $\dot{\mathcal{V}}$ is simplified to 
\begin{align}  \label{lyap_case2}
\dot{\mathcal{V}} \leq -\varrho \mathcal{V} + \varrho \frac{\bar{\zeta}}{\underline{\zeta}}  +  \frac{1}{2}\sum \limits_{i=0}^{3} \nu_{i} {\mathcal{H}^*_{i}}^2 + \frac{{{\epsilon}}}{\underline{\zeta}}
 \end{align}
By comparing \eqref{lyap_case1} and \eqref{lyap_case2}, the combined derivative of ${\mathcal{V}}$ for both Case (i) and Case (II), can be bounded as 
\begin{align} \label{lyap_final}
\dot{\mathcal{V}}  \leq& -\varrho \mathcal{V} + \delta, \\
\text{where}~ \delta =&  \varrho \frac{\bar{\zeta}}{\underline{\zeta}}  +  \frac{1}{2}\sum \limits_{i=0}^{3} \nu_{i} {\mathcal{H}^*_{i}}^2 + \frac{{{\epsilon}}}{\underline{\zeta}}  \nonumber
\end{align}
Defining a scalar $\kappa$ as $0<\kappa<\varrho$, (\ref{lyap_final}) can be simplified to
\begin{align}
\dot{\mathcal{V}} & \leq  -\kappa \mathcal{V} - (\varrho - \kappa)\mathcal{V} + \delta.
\end{align}
Further defining a scalar $\mathcal{  B} = \frac{\delta}{(\varrho - \kappa)}$
 it can be concluded that $\dot{\mathcal{V}} (t) < - \kappa \mathcal{V} (t)$ when $\mathcal{V} (t) \geq \mathcal{ B}$, so that
\begin{align}
    \mathcal{V} & \leq \max \{ \mathcal{V}(0), \mathcal{  B} \}, ~\forall t \geq 0, \label{final}
\end{align}
and the closed-loop system remains UUB (cf. UUB definition $4.6$ as in \cite{khalil2002nonlinear}). This guarantees boundedness of $\mathbf{s}$, implying boundedness of $\Delta \mathcal{I}$ from (\ref{I_cl}) (using (\ref{gain_new})). This further implies boundedness of $\mathbf{e}, \dot{\mathbf{e}}$ from (\ref{deviation_error}).

\begin{remark}[Role of \(\zeta\)]
Note that \(\|\mathbf{s}\| < \varpi\) does not automatically imply that \(\|\ddot{\boldsymbol{\chi}}\|\) or \(\|\boldsymbol{\xi}\|\) remain bounded. To obtain boundedness, the auxiliary gain \(\zeta\), adapted via \(\dot{\zeta}\), cancels the cross terms (e.g., \(\hat{\mathcal H}_{3}\|\ddot{\boldsymbol{\chi}}\|\|\mathbf{s}\|\)) to ensure that \(\dot{\mathcal{V}}\) remains negative outside a bounded region.
\end{remark}

\bibliographystyle{IEEEtran}

\begin{thebibliography}{10}
\providecommand{\url}[1]{#1}
\csname url@samestyle\endcsname
\providecommand{\newblock}{\relax}
\providecommand{\bibinfo}[2]{#2}
\providecommand{\BIBentrySTDinterwordspacing}{\spaceskip=0pt\relax}
\providecommand{\BIBentryALTinterwordstretchfactor}{4}
\providecommand{\BIBentryALTinterwordspacing}{\spaceskip=\fontdimen2\font plus
\BIBentryALTinterwordstretchfactor\fontdimen3\font minus \fontdimen4\font\relax}
\providecommand{\BIBforeignlanguage}[2]{{%
\expandafter\ifx\csname l@#1\endcsname\relax
\typeout{** WARNING: IEEEtran.bst: No hyphenation pattern has been}%
\typeout{** loaded for the language `#1'. Using the pattern for}%
\typeout{** the default language instead.}%
\else
\language=\csname l@#1\endcsname
\fi
#2}}
\providecommand{\BIBdecl}{\relax}
\BIBdecl

\bibitem{kim_iros13}
S.~Kim, S.~Choi, and H.~J. Kim, ``Aerial manipulation using a quadrotor with a two dof robotic arm,'' in \emph{2013 IEEE/RSJ International Conference on Intelligent Robots and Systems}, 2013, pp. 4990--4995.

\bibitem{heredia_iros14}
G.~Heredia, A.~Jimenez-Cano, I.~Sanchez, D.~Llorente, V.~Vega, J.~Braga, J.~Acosta, and A.~Ollero, ``Control of a multirotor outdoor aerial manipulator,'' in \emph{2014 IEEE/RSJ International Conference on Intelligent Robots and Systems}, 2014, pp. 3417--3422.

\bibitem{Lippiello_12}
V.~Lippiello and F.~Ruggiero, ``Exploiting redundancy in {Cartesian} impedance control of {UAVs} equipped with a robotic arm,'' in \emph{2012 IEEE/RSJ International Conference on Intelligent Robots and Systems}, 2012, pp. 3768--3773.

\bibitem{ollero_15}
M.~I. SÃ¡nchez, J.~Ã. Acosta, and A.~Ollero, ``Integral action in first-order closed-loop inverse kinematics. application to aerial manipulators,'' in \emph{2015 IEEE International Conference on Robotics and Automation (ICRA)}, 2015, pp. 5297--5302.

\bibitem{Cataldi_16}
E.~Cataldi, G.~Muscio, M.~A. Trujillo, Y.~Rodriguez, F.~Pierri, G.~Antonelli, F.~Caccavale, A.~Viguria, S.~Chiaverini, and A.~Ollero, ``Impedance control of an aerial-manipulator: Preliminary results,'' in \emph{2016 IEEE/RSJ International Conference on Intelligent Robots and Systems (IROS)}, 2016, pp. 3848--3853.

\bibitem{rashad_energy_2022}
R.~Rashad, D.~Bicego, J.~Zult, S.~Sanchez-Escalonilla, R.~Jiao, A.~Franchi, and S.~Stramigioli, ``\BIBforeignlanguage{en}{Energy {aware} {impedance} {control} of a {flying} {end}-{effector} in the {port}-{Hamiltonian} {framework}},'' \emph{\BIBforeignlanguage{en}{IEEE Transactions on Robotics}}, vol.~38, no.~6, pp. 3936--3955, Dec. 2022.

\bibitem{Zhang_TASE_24}
Z.~Zhang, Y.~Chen, Y.~Wu, B.~He, Z.~Miao, H.~Zhang, and Y.~Wang, ``Hybrid force/position control for switchable unmanned aerial manipulator between free flight and contact operation,'' \emph{IEEE Transactions on Automation Science and Engineering}, vol.~21, no.~3, pp. 4283--4297, 2024.

\bibitem{guanya}
X.~Guo, G.~He, J.~Xu, M.~Mousaei, J.~Geng, S.~Scherer, and G.~Shi, ``Flying calligrapher: Contact-aware motion and force planning and control for aerial manipulation,'' \emph{IEEE Robotics and Automation Letters}, vol.~9, no.~12, pp. 11\,194--11\,201, Dec 2024.

\bibitem{7989314}
H.~W. Wopereis, J.~J. Hoekstra, T.~H. Post, G.~A. Folkertsma, S.~Stramigioli, and M.~Fumagalli, ``Application of substantial and sustained force to vertical surfaces using a quadrotor,'' in \emph{2017 IEEE International Conference on Robotics and Automation (ICRA)}, 2017, pp. 2704--2709.

\bibitem{suarez2018physical}
A.~Suarez, G.~Heredia, and A.~Ollero, ``Physical-virtual impedance control in ultralightweight and compliant dual-arm aerial manipulators,'' \emph{IEEE Robotics and Automation Letters}, vol.~3, no.~3, pp. 2553--2560, 2018.

\bibitem{tognon2019truly}
M.~Tognon, H.~A.~T. Ch{\'a}vez, E.~Gasparin, Q.~Sabl{\'e}, D.~Bicego, A.~Mallet, M.~Lany, G.~Santi, B.~Revaz, J.~Cort{\'e}s \emph{et~al.}, ``A truly-redundant aerial manipulator system with application to push-and-slide inspection in industrial plants,'' \emph{IEEE Robotics and Automation Letters}, vol.~4, no.~2, pp. 1846--1851, 2019.

\bibitem{wang_impact_2024}
S.~Wang, Z.~Ma, F.~Quan, and H.~Chen, ``\BIBforeignlanguage{en}{Impact {absorbing} and {compensation} for {heavy} {object} {catching} {with} an {unmanned} {aerial} {manipulator}},'' \emph{\BIBforeignlanguage{en}{IEEE Robotics and Automation Letters}}, vol.~9, no.~4, pp. 3656--3663, Apr. 2024.

\bibitem{orsag2018aerial}
M.~Orsag, C.~Korpela, P.~Oh, S.~Bogdan, and A.~Ollero, \emph{Aerial manipulation}.\hskip 1em plus 0.5em minus 0.4em\relax Springer, 2018.

\bibitem{arleo2013control}
G.~Arleo, F.~Caccavale, G.~Muscio, and F.~Pierri, ``Control of quadrotor aerial vehicles equipped with a robotic arm,'' in \emph{21St Mediterranean Conference on Control and Automation}.\hskip 1em plus 0.5em minus 0.4em\relax IEEE, 2013, pp. 1174--1180.

\bibitem{spong2020robot}
M.~W. Spong, S.~Hutchinson, and M.~Vidyasagar, \emph{Robot Modeling and Control}.\hskip 1em plus 0.5em minus 0.4em\relax Wiley, 2020.

\bibitem{kim2013aerial}
S.~Kim, S.~Choi, and H.~J. Kim, ``Aerial manipulation using a quadrotor with a two dof robotic arm,'' in \emph{2013 IEEE/RSJ International Conference on Intelligent Robots and Systems}.\hskip 1em plus 0.5em minus 0.4em\relax IEEE, 2013, pp. 4990--4995.

\bibitem{Choosing_impedance_parameters}
M.~J. Pollayil, F.~Angelini, G.~Xin, M.~Mistry, S.~Vijayakumar, A.~Bicchi, and M.~Garabini, ``Choosing stiffness and damping for optimal impedance planning,'' \emph{IEEE Transactions on Robotics}, vol.~39, no.~2, pp. 1281--1300, 2023.

\bibitem{billard16_impedance}
K.~Kronander and A.~Billard, ``Stability considerations for variable impedance control,'' \emph{IEEE Transactions on Robotics}, vol.~32, no.~5, pp. 1298--1305, 2016.

\bibitem{richter2016polynomial}
C.~Richter, A.~Bry, and N.~Roy, ``Polynomial trajectory planning for aggressive quadrotor flight in dense indoor environments,'' in \emph{Robotics Research}.\hskip 1em plus 0.5em minus 0.4em\relax Springer, 2016, pp. 649--666.

\bibitem{berscheid2021jerk}
L.~Berscheid and T.~Kr{\"o}ger, ``Jerk-limited real-time trajectory generation with arbitrary target states,'' \emph{ArXiv}, vol. abs/2105.04830, 2021.

\bibitem{khalil2002nonlinear}
H.~K. Khalil, \emph{Nonlinear Systems}, ser. Pearson Education.\hskip 1em plus 0.5em minus 0.4em\relax Prentice Hall, 2002.

\end{thebibliography}
% Generated by IEEEtran.bst, version: 1.14 (2015/08/26)

\nomenclature{\(b_r\)}{Width of right side module}

\end{document}